\documentclass[11pt]{article}

\usepackage{a4}

\usepackage{amsfonts}

\usepackage{cite}

\newcommand{\pref}[1]{(\ref{#1})}
\newcommand{\nn}{\nonumber}

\newcommand {\bc} {\begin{center}}
\newcommand {\ec} {\end{center}}

\newcommand {\be} {\begin{equation}}
\newcommand {\ee} {\end{equation}}                                         
\newcommand {\bea} {\begin{eqnarray}}
\newcommand {\eea} {\end{eqnarray}}                                         
\newcommand {\ba} {\begin{array}}
\newcommand {\ea} {\end{array}}

\newcommand{\tr}{\mathop{\mathrm{tr}}}
\newcommand{\Tr}{\mathop{\mathrm{Tr}}}

\newcommand{\gdg}[1]{\Omega \partial_{#1} \Omega^{-1}}
\newcommand{\dpar}[1]{\partial_{#1}}

\newcommand{\id}{{\rm 1\kern-.12em
\rule{0.3pt}{1.5ex}\raisebox{0.0ex}{\rule{0.1em}{0.3pt}}}\,}

\newcommand {\dL} {\partial_{t} \Lambda_{t}}

\newcommand \Lt {\Lambda_{t}}
\def\Lti {\Lambda^{-1}_{t}}
\def\Pt  {P_{t}}
\def\PO  {P_0}
\def\Qt  {Q_{t}}

\def\Xt  {X_{t}}
\def\MT  {{\mathcal T}}

\def\psibar{\overline{\psi}}

\newcommand{\twist}{\mbox{${\cal T}$}}
\newcommand{\dsymR}{\mbox{$d_{R}^{abc}$}}
\newcommand{\dsym}{\mbox{$d^{abc}$}}

\newsymbol\blacksquare 1004

\newsymbol\newsquare 1003

\begin{document}

\title{
\begin{flushleft}
{\small
\hfill {\sc DESY}-99/188}
\end{flushleft}
\vspace{2cm}
Global Anomalies in\\
Chiral Gauge Theories on the Lattice}

\author{Oliver B\"ar and Isabel Campos}
\maketitle

\begin{center}
{\small {\sl Deutsches Elektronen-Synchrotron}, \\
 {\rm Notkestrasse 85, 22603 Hamburg (Germany)} \\
 e-mail: \tt obaer@mail.desy.de, icampos@mail.desy.de} \\
\end{center}

\bigskip

\begin{abstract}
We discuss the issue of global anomalies in chiral gauge
theories on the lattice. In L\"{u}scher's approach, these obstructions
make it impossible to 
define consistently a fermionic measure for the path integral. 
We show that an $SU(2)$ theory has such a global anomaly if the Weyl
fermion is in the fundamental representation.
The anomaly in higher representations is also discussed. We finally
show that this obstruction is the lattice analogue of the $SU(2)$ 
anomaly first discovered by Witten.

\end{abstract}

\newpage

%
%

\setcounter{equation}{0}
\renewcommand{\theequation}{1.\arabic{equation}}

\vspace*{0.5cm}
\noindent {\large {{\bf {{\slshape {1. Introduction}}}}}}
\vspace*{0.5cm}

\noindent The formulation of chiral lattice gauge theories
on firm theoretical grounds has recently turned out to be an interesting
and rapidly developing field \cite{PisaReview}.
At the root of the recent achievements is the realization that the 
Nielsen-Ninomiya theorem \cite{NN}
can be circumvented whenever the lattice Dirac operator
fulfils the Ginsparg-Wilson relation \cite{GinspargWilson}
\bea
D \gamma_5 + \gamma_5 D = a D \gamma_5 D \ .
\label{gw}
\eea
Lattice actions defined using these kind of operators
exhibit good chiral properties \cite{Penhiscola,DomainWall,Overlap}.
In particular, L\"uscher has recently proved that within this framework
chiral theories with manifest gauge invariance can be
defined on the lattice without compromising the theory in any
other way \cite{AbelianChGT,NonAbelianChGT}. 

In this work we apply this formalism to the study of global anomalies
on the lattice.
In the continuum theory global anomalies were first discovered by
Witten \cite{Witten}, who proved the mathematical inconsistency of the 
$SU(2)$ gauge theory coupled 
to an odd number of doublets of Weyl fermions. Let us briefly sketch
his argumentation.

Consider an $SU(2)$ gauge theory 
coupled to a single doublet of massless Dirac fermions.
In terms of the gauge potential $A_{\mu}$ and the fermion fields
$\psi,\psibar$ the action reads  
\bea
S(A_{\mu},\psi,\psibar) = S_g(A_{\mu}) + \int \mbox{d}^4 x \: \psibar D \psi .
\label{cont_action}
\eea
$S_g$ denotes the pure gauge part of the action and $D$ stands for
the hermitian Dirac operator. In the vacuum sector,
after performing the fermion integration, we obtain the
following partition function
\bea
{\cal Z}_{\rm {Dirac}} = \int  {\mathcal D}[A] {\mathcal D}[\psi] 
{\mathcal D}[\psibar] \: 
{\rm e}^{-S(A,\psi,\psibar)}
= \int {\mathcal D}[A] \: \det D(A) \: {\rm e}^{-S_g(A)} .
\eea
In $SU(2)$, the fermion determinant is real. In addition,
one doublet of Dirac fermions is equivalent to two 
doublets of left-handed Weyl fermions. Therefore having only one doublet of
Weyl fermions the partition function is given by
\bea
{\cal Z}_{\rm {Weyl}} = \int  {\cal D}[A] {\cal D}[\psi] 
{\cal D}[\psibar] \: 
{\rm e}^{-S(A,\psi,\psibar)} 
= \int {\cal D}[A] \: \sqrt{\det D(A)} \: {\rm
e}^{-S_g(A)} \ .  
\label{Z_weyl}
\eea
Here the sign of the square root is undetermined. Fixing the sign is
therefore part of the complete definition of the chiral gauge
theory. Let us arbitrarily fix the sign for the classical vacuum
configuration.  
If we insist on a smooth gauge field dependence  of
$\sqrt{\det D(A)}$, the sign is also fixed for all gauge fields
in the vacuum sector.
However, using this prescription for the sign the square root of the
Dirac operator is not gauge invariant.
Witten has shown this as follows.

One first notices that in four dimensions there exist 
$SU(2)$ gauge transformations
that cannot be continuously deformed to the identity mapping.
In mathematical terms this  is expressed by saying that the fourth
homotopy group of $SU(2)$ is non-trivial: 
$\pi_4[SU(2)] ={\mathbb {Z}}_2$. 

Now take any gauge 
transformation $g(x)$ in the non-trivial
homotopy class of $SU(2)$ and consider the path 
\bea
A_{\mu}^t = tA_{\mu} + (1-t) A_{\mu}^g  \ , \quad  t \in [0,1] \ ,  
\label{path}
\eea
in the vector space of gauge potentials. This path
is well defined and interpolates smoothly 
between $A_{\mu}$ and its gauge transform
\be
A_{\mu}^g = g (A_{\mu} + \partial_{\mu}) g^{-1} .
\ee
Using the Atiyah-Singer index theorem Witten proved that  
an odd number of eigenvalues of the square root of the Dirac operator
crosses zero as a function of $t$. This leads to the switch of sign 
\bea
\sqrt{\det D(A)} = - \sqrt{\det D(A^g)}.
\eea
Hence gauge invariance is violated.

On the other hand we could define the sign always positive, taking the
absolute value of the square root of $\det D(A)$.
In that way gauge invariance is achieved but the smoothness of
$\sqrt{\det D(A)}$ has been sacrificed.
The square root of the fermion determinant 
would have a cusp for the $t$ value
in which the crossing of the eigenvalues takes place.
In conclusion, Witten's anomaly simply states that the sign of the
square root for $\det D(A)$ cannot be defined in a smooth and
gauge invariant way. 

As has become clear, global anomalies are related
to the properties of the space of gauge orbits.
In perturbation theory the gauge fixing procedure makes it impossible
that the gauge field goes along full orbits and prevents the 
observation of global obstructions. This is the reason why
global anomalies are referred to as non-perturbative.

The formulation of chiral lattice gauge theories recently
proposed by L\"uscher, besides being non-perturbative, 
is manifestly gauge invariant. 
As a consistency requirement of the whole approach, 
both perturbative and non-perturbative
continuum anomalies should cancel in the lattice 
formulation. 
Concerning the perturbative ones the status is 
satisfactory from a theoretical point of view. Only those theories
that are perturbatively anomaly free in the continuum can be consistently
formulated on the lattice \cite{NonAbelianChGT}. 
Now it is natural to ask whether the known
non-perturbative anomalies are also reproduced by the lattice theory.
The aim of this work is to clarify this issue. 

The paper is organized in the following way.
We first define Weyl fermions on the lattice and discuss the gauge
field dependence of the corresponding projection operators (section
2), a new feature due to the Ginsparg-Wilson relation.
In section 3, following reference \cite{NonAbelianChGT}, 
we review the definition of a fermion measure for chiral
lattice gauge theories and define what is meant by a global anomaly.
In section 4 we investigate the $SU(2)$ gauge theory
with a single doublet of Weyl fermions. We will find that this theory
shows a global anomaly in
the sense previously defined. The anomaly for higher
$SU(2)$ representations is discussed in section 5.
Finally we show that this 
anomaly is equivalent to the Witten anomaly  in the
continuum (section 6).
The last section is devoted to conclusions
and the most technical derivations are deferred to Appendix A.

This work is intimately connected to ref. \cite{NonAbelianChGT} to
which the reader is referred for full explanations on some results
used here. We also carry over completely the notations used there.

%
%

\setcounter{equation}{0}
\renewcommand{\theequation}{2.\arabic{equation}}

\vspace*{0.5cm}
\noindent {\large {\bf { {\slshape{2. Weyl fermions on the lattice}}}}}
\vspace*{0.5cm}

\noindent {\bf {\slshape {2.1 Preliminaries}}}
\vspace*{0.5cm}

\noindent We consider space-time to be restricted
to the sites of a finite euclidean lattice with periodic
boundary conditions. Gauge fields are specified by
group elements on the bonds joining consecutive sites. 
Fermionic fields are represented by variables defined
on the lattice sites.
In the following we consider that the gauge field 
couples to a multiplet of left-handed fermions given in
some unitary representation $R$ of the gauge group.

As has been shown recently \cite{AbelianChGT,NonAbelianChGT}
a proper definition of these lattice theories
is possible if the Dirac operator satisfies the Ginsparg-Wilson
relation. An explicit example has been proposed by Neuberger
\cite{Neuberger_operator}. However, the locality and differentiability 
with respect to
the gauge field of this operator is only guaranteed if the gauge field
satisfies the bound
\be\label{bound_plaquette}
\|1-R [U(p)] \|\,\leq\,\frac{1}{30}
\ee
for all plaquettes $U(p)$ \cite{LuscherLocality}. Fields
compatible with this bound 
are called {\em admissible} and we restrict ourselves to those.

The Ginsparg-Wilson relation leads to  exact chiral symmetry on the
lattice\cite{ExactSymmetry}. The fermion action splits into a left-
and a right-handed part if we introduce the chiral projectors
\cite{BoulderReview,OverlapSplit} 
\bea
\hat{P}_{\pm} = \frac{1}{2} [1 \pm  \gamma_5 (1 - a D)]\ ,
\quad P_{\pm} = \frac{1}{2} (1 \pm \gamma_5) \ .
\label{projectors}
\eea
In particular, left-handed fields are defined by the conditions
\bea
\label{constraint1}
\hat{P}_- \psi &=& \psi  \ ,\\
\psibar P_+ &=& \psibar \ .
\label{constraint2}
\eea
The fermionic part of the lattice action describing left-handed (Weyl)
fermions only therefore reads 
\bea
S_{F,L} & = & a^4\sum_x \psibar(x)[P_+D\hat{P}_- \psi](x)\,.
\eea
Note that $\hat{P}_-$ contains the Dirac operator $D$.
The definition of left-handed fields therefore depends on the
particular background gauge field.
This has some geometrical implications 
which will be relevant later for our investigation
of global anomalies.

%
%

\vspace*{0.5cm}
\noindent {\bf {\slshape {2.2 Paths in configuration space}}}
\vspace*{0.5cm}

\noindent 
We will be interested in trajectories in configuration space.
For $t \in [0,1]$ a trajectory or path
$\Gamma$ is defined by specifying
the configuration of the gauge field, $U_t(x,\mu)$, at each 
value of the parameter $t$. 

We require the gauge field to be a smooth function
of $t$. 
Along paths in configuration space we will use
the notation $\Pt \equiv \hat{P}_-|_{U=U_t}$ 
to refer to $\hat{P}_-$ along the path. 
Eq. \pref{constraint1} then implies that
the subspace of left-handed fields does not remain
constant but rather rotates 
according to the evolution of $P_t$. We denote by $\Qt$  
the operator transporting $\Pt$ along the path. This operator
is defined to be the solution of the differential equation
\bea
\dpar{t} \Qt = [\dpar{t} \Pt,\Pt] \Qt \ , \qquad\mbox{$Q_0 = \id$} .
\label{DE}
\eea
The hermiticity of $\Pt$ implies that $\Qt$ is an unitary operator.
That $\Qt$ is indeed the transporter of $\Pt$ can be seen 
by verifying the relation 
\bea
\Pt \Qt & = & \Qt P_0.
\eea
Therefore if
$\psi_0$ is a left-handed field at the point $t=0$, at a different
point $t$ we have
$  \Pt \, \Qt \, \psi_0 
 \,= \,\Qt \, \psi_0$.
This means that $\Qt \,\psi_0$ is a left-handed field at 
the point $t$.

The definition of $\Qt$ implicitly assumes that the gauge
field $U_t (x,\mu)$ is differentiable 
for all $t$. However, it can be extended straightforwardly 
to piecewise smooth paths.
For that one defines $Q_t^k$ using eq. \pref{DE} for each
individual smooth piece $\Gamma^k$. The operator
$\Qt$ along the whole path is then the ordered product
\bea
Q_t = \prod_{k=1}^n Q_t^k \ .
\label{Q_piece}
\eea
This definition is consistent, that is whenever we split a path in
more pieces than necessary we will obtain the same result for $Q_t$.
The unitarity of $\Qt$
follows from the unitarity of the individual $Q_t^k$.

\vspace*{0.5cm}
\noindent {\bf {\slshape {2.3 Definition of the twist}}}
\vspace*{0.5cm}

\noindent
Consider now closed loops in configuration
space, that is $U_0 (x,\mu)=U_1 (x,\mu)$.
After a whole turn the subspace of left-handed fields has rotated 
following the gauge field geometry along the loop.
A priori there is no reason to expect the operator $Q_1$ to be
the identity mapping.

As a measure for this rotation we introduce the $twist$ $\MT$ for any
closed loop $\Gamma$ in configuration space, defined by 
\bea
\MT \, (\Gamma) = \det \, [1 - P_0 + P_0 \, Q_1 ] \ .
\label{def_twist}
\eea
Note that the definition of the determinant includes the 
operator $(1 - P_0)$, representing the identity
in the space of right-handed fields at $t=0$. This operator ensures
a well defined determinant in the whole space
of fermion fields.

The unitarity of $\Qt$ implies that $\MT$ is phase for all gauge
groups. If $Q_1$ is the identity, we get trivially a twist
equal to one. 
The twist for two closed loops with the same base point satisfies
the composition law
\bea
\MT \, (\Gamma_1 \circ \Gamma_2) = \MT \, (\Gamma_1) \cdot \MT \,
(\Gamma_2) \ .
\label{composition_law}
\eea
In the following the twist is called non-trivial when it is different
from one. 
Closed loops with non-trivial twist will play an important 
role for the discussion of global
anomalies throughout the next sections.

\setcounter{equation}{0}
\renewcommand{\theequation}{3.\arabic{equation}}

\vspace*{0.5cm}

\noindent{\large{{\bf{{\slshape{3. Fermion integration measure
and global anomalies}}}}}}
\vspace*{0.5cm}
  
\noindent{\bf{{\slshape{3.1 Definition of the fermion measure}}}}
\vspace*{0.5cm}

\noindent 
To set up a {\em quantum} field theory we need to define a
functional integral in order to compute correlation functions.
In the present approach to chiral gauge theories
the projector $\hat{P}_-$ depends on the gauge field 
[cf. eq. \pref{constraint1}]. This implies that the fermion
measure for left-handed fields is also gauge-field-dependent.
Choosing a measure valid for all gauge field configurations 
is therefore a non-trivial task.
On the other hand, the cancellation of the gauge anomaly and the locality
of the resulting theory are fundamental properties that one needs to
preserve when defining the fermion integration measure.

In \cite{NonAbelianChGT} it is shown  that the
definition of the fermion measure is equivalent to finding a 
Lie algebra valued current $j_{\mu}(x) = j_{\mu}^c (x) \, T^c$,
where $c$ denotes the colour index,
satisfying various conditions:\newline

\noindent {\em 1. Smoothness condition:} $j_{\mu}(x)$ has to be a
  smooth function 
  of the gauge field.\newline 

\noindent {\em 2. Locality condition:} $j_{\mu}(x)$ has to be a
  local expression in the gauge field.\newline

\noindent {\em 3. Gauge invariance condition:} The current is
gauge covariant and satisfies the 
anomalous conservation law
\be\label{gauge_inv_condition}
\big(\nabla_{\!\mu}^*j_{\mu}\big)^c(x)  = {\cal A}^c
(x),\label{covariance_condition}
\ee
\be
{\cal A}^c (x)  =  \frac{ia}{2}\mbox{tr}\left\{\gamma_5
  R(T^c)D(x,x)\right\}\label{def_cov_anomaly} \ ,
\ee

\noindent where  $R(T^c)$ denotes the generator of the gauge group,
$\nabla_{\!\mu}^*$ is the gauge covariant backward lattice derivative
and $D(x,x)$ is the kernel of the Dirac operator.
The expression ${\cal A}^c (x)$ represents the covariant anomaly on
the lattice. \newline

\noindent The formulation of the fourth and last condition requires some 
definitions. For any smooth path $U_t(x,\mu)$, $0\leq t\leq 1$, we
define the {\em Wilson line}
\be\label{def_Wilson_line}
W=\exp\left\{i\int_0^1 \mbox{d}t {\mathcal L}_\eta\right\},
\ee
where the so-called {\em measure term} $\mathfrak{L}_{\eta}$ is defined by
\be
\mathfrak{L}_{\eta}\, =\, a^4\sum_x \eta_{\mu}^c(x)j_{\mu}^c(x),\qquad
a\eta_{\mu}(x) = \partial_t U_t(x,\mu)U_t(x,\mu)^{-1}. 
\label{deformation}
\ee
The Wilson line depends on the  
chosen current $j_{\mu}^c(x)$ and will be the total change of phase of
the reconstructed fermion measure along the path.\newline

\noindent {\em 4. Integrability condition:} $j_{\mu}^c(x)$ has to satisfy 
\be\label{definition_integrability_condition}
W={\cal T}
\ee
for all closed curves in configuration space.  The twist ${\cal T}$
is defined in (\ref{def_twist}).\newline

\noindent Whenever a current meets all four  conditions, a fermion measure 
can be consistently defined. The chiral lattice gauge theory is then -- up to
a constant phase in the path integral -- fully specified. 
This result sometimes goes under the name  
{\em Reconstruction Theorem} and the details 
can be found in \cite{NonAbelianChGT}. 
We want to emphasize that neither the
locality of the theory is
violated nor is gauge invariance broken within this framework.

The first condition guarantees that the reconstructed
fermion measure depends smoothly on the gauge field.
The locality of the entire theory is preserved as long as condition 2 holds.
Once the third condition is fulfilled,
the fermion measure is gauge invariant for infinitesimal gauge
transformations. 
Finally, the measure is path-independent if and
only if the integrability condition holds.

We remark that the conditions are only sensitive to the
axial vector part of the current. Once we have found a current
compatible with the conditions, we can easily project out the axial
vector part of it. This will also satisfy the
conditions. Hence without loss of generality we ask for a current that
transforms as an axial vector under the lattice symmetries.
\vspace*{0.5cm}

\noindent {{{\bf{{\slshape{3.2 Global anomalies
}}}}}}
\vspace*{0.5cm}

\noindent It is by no means clear that for a given gauge group and a fixed
lattice spacing $a$ an appropriate current $j_{\mu}(x)$ exists which
satisfies all four conditions. Even if we know the existence it is
fairly difficult to construct it. So far the existence of a current has
been rigorously proved only for the abelian gauge group $U(1)$
\cite{AbelianChGT}. The proof for the non-abelian case is still
missing.

In the following we speak of a {\em global anomaly} if any current
satisfying the first three conditions necessarily violates the last one.
In that case no proper fermion measure exists.
Such an anomaly is therefore an insurmountable obstruction in 
formulating a chiral lattice gauge theory. The reason why we call this
a global anomaly will become clear later.

In a semi-classical analysis global anomalies arise in the
following way. Consider classical fields,
originating from some smooth gauge potential $A_{\mu}(x)$ by the
path-ordered exponential 
\be\label{classical_field}
U(x,\mu) = {\cal P} \exp\left\{ \int_0^1\mbox{d}t
A_{\mu}(x+(1-t)a\hat{\mu})\right\}. 
\ee
If the current $j_{\mu}^c$ is a local and smooth function of the
gauge field, as required by the first two conditions, it can be
expanded in powers of the lattice spacing $a$.
Moreover, if the current satisfies the gauge invariance condition,
$j_{\mu}^c$ is a 
local gauge covariant polynomial of dimension 3 in the gauge potential that
transforms as an axial vector. Therefore the
expansion in the lattice spacing starts linearly in $a$, i.e.
\be\label{asymptotic_current}
j_{\mu}^c(x) = 0 + {\cal O}(a).
\ee
So far we employed the first three conditions only.
Any current compatible with them vanishes in the classical continuum
limit.
This immediately implies 
\be\label{asymptotic_Wilsonline}
W = 1+{\cal O}(a)
\ee 
for the Wilson line (\ref{def_Wilson_line}) along a smooth curve of
classical fields. 
This can lead to a conflict with the integrability
condition (\ref{definition_integrability_condition}) if the twist
along such a  closed curve is non-trivial in the continuum limit.
In that case, any current satisfying the first three conditions
will necessarily violate the integrability condition for some small
enough lattice spacing $a$. Consequently, a proper fermion measure cannot 
be defined.

%
%
\setcounter{equation}{0}
\renewcommand{\theequation}{4.\arabic{equation}}

\vspace*{0.5cm}
\noindent {\large { {\bf {{\slshape {4. Global anomaly in ${\bf SU(2)}$}}}}}}
\vspace*{0.5cm}

\noindent 
We now consider the special case when the gauge field is in the 
fundamental representation ($j=1/2$) of $SU(2)$. 
It will be shown that 
there exist loops in configuration space with a non-trivial 
twist. Higher $SU(2)$ representations are discussed in section 5.

\vspace*{0.5cm}
\noindent {\bf {\slshape {4.1 Definition of closed loops}}}
\vspace*{0.5cm}

\noindent Due to the reality properties of $SU(2)$
the twist is real and therefore equal to $\pm 1$.
Since these values cannot change continously
the twist is a homotopy invariant for $SU(2)$.
Concerning global anomalies,
violations of the integrability condition will
manifest themselves as closed loops in configuration space with 
a twist equal to $-1$.
In the following we will explicitly construct one such loop.
 
%
%

To begin with we take a 
homotopically non-trivial continuum gauge transformation $g(x)$.
The particular form of $g(x)$ is not needed.
Restricting the space-time points to the sites of an euclidean lattice
defines a lattice gauge transformation.
For $t \in [0,1]$ we define two paths in configuration space
\bea
\Gamma_1: & & U_t (x,\mu) = g(x)^t g(x + a\mu)^{-t} \label{def_path_1} \\[1ex]
\Gamma_2: & & U_t (x,\mu) = [g(x) g(x + a\mu)^{-1}]^{1-t}\label{def_path_2} 
\label{paths}
\eea
On the lattice all gauge transformations are smoothly connected
with the identity mapping. In particular,
$\Gamma_1$ is a smooth curve made up of $SU(2)$ gauge transformations.
This path is a pure lattice artifact and has no analogue in the
continuum.  
$\Gamma_2$ also connects the vacuum configuration  with its gauge transformed
and is just the lattice version of the path \pref{path}.

Consider the closed loop $\Gamma_2 \, \circ \, \Gamma_1$. It
starts and ends at the classical
vacuum passing through the pure gauge configuration $g(x)g(x+a\mu)^{-1}$.
Our aim is computing $\MT (\Gamma_2 \,\circ \, \Gamma_1)$. 

As an auxiliary tool for the calculation we
define a third path. 
First note that
$SU(2)$ can be embedded in a group with
a trivial fourth homotopy group $\pi_4$, $SU(3)$ say.
The vacuum configuration  and its
gauge transformed by $g(x)$ are  
connected by a smooth 
path of gauge transformations in $SU(3)$. To be more precise there
exists a path $\Omega(s,x)$, $0\leq s \leq 1$, with 
boundary values \cite{Embedding}
\be\label{SU3_path}
\Omega(0,x) \,=\,\id\,, \qquad \Omega(1,x) \, = \,\left(
\begin{array}{c|c}
g(x)&\rule[-3.5mm]{0mm}{8mm}\\\hline
 & 1
\end{array}\right)\,.
\ee
Here the fundamental representation of $SU(3)$ has been considered. Having this
$\Omega$ at hand we define a third path on the lattice by
\be\label{def_gamma_3}
\Gamma_3:  \quad U_s (x,\mu) = \Omega(s,x) \Omega(s,x + a\mu)^{-1},\nonumber
\ee
which connects the vacuum with the 
pure gauge configuration $g(x)g(x+a\mu)^{-1}$
in the $1/2 \,\oplus \, 0$ representation of $SU(2)$.
If we enlarge the first two paths
also to this representation we can use the composition law 
(\ref{composition_law}),
\bea
\MT(\Gamma_2\circ\Gamma_1) & =
&\MT(\Gamma_2\circ\Gamma_3)\cdot\MT(-\Gamma_3\circ\Gamma_1),
\label{compo}
\eea
to compute the twist along $\Gamma_2\circ\Gamma_1$. The additional
singlet representation does not affect our final result. To see this
one first notices the relation
\bea\label{direct_sum_twist}
\MT(\Gamma_{j_1\oplus j_2}) & = &
\MT(\Gamma_{j_1})\cdot\MT(\Gamma_{j_2}),
\eea
that holds for a direct sum of two $SU(2)$ representations. In
addition the twist  
is equal to 1 for the singlet representation. Therefore the left-hand
side of \pref{compo} is also equal to the twist in the fundamental
representation alone.

Two comments should be made.
We want to emphasize that the path $\Gamma_2$ is well-defined in the
space of admissible fields provided that the lattice spacing $a$ is
small enough. In the following we always assume this to be the case.
The fields along $\Gamma_1$ and $\Gamma_3$ trivially
satisfy the bound \pref{bound_plaquette} because they are pure gauge
transformations. 

The paths $\Gamma_i$, $i=1,2,3$, are
smooth. However, the loops we have defined out of them
are only piecewise smooth.
At the contact points of the paths in the vacuum and the pure gauge
configuration $g(x)g(x + a\mu)^{-1}$ the loops are not differentiable
with respect to $t$.

%
%

\vspace*{0.5cm}
\noindent{\bf {\slshape {4.2 Twist along closed gauge loops}}}
\vspace*{0.5cm}

\noindent 
A closed expression for the twist can be derived for pure gauge
loops.
The discussion of this section applies to any gauge group, only
in the end we will restrict ourselves to our particular
gauge loop in $SU(3)$.

Given an initial gauge field $U_0(x,\mu)$ gauge 
paths in configuration space are of the form
\bea
U_t(x,\mu) = \Lt(x) \: U_0(x,\mu) \: \Lt(x + a\mu)^{-1} \ ,
\eea
where $\Lt$ denotes a curve of gauge transformations with
initial value $\Lambda_0 = \id$. The projector to the subspace of
left-handed fermion fields along gauge paths is thus given by
\bea
\Pt = \Lt P_0 \Lti \ .
\eea
For gauge paths the differential equation (\ref{DE}) for the evolution
operator $Q_t$ can be solved.
In terms of $\Xt = \Lti \dL$ the solution reads
\bea
\Qt = \Lt \, {\cal P} 
\exp \left( - \int_0^t \: \mbox{d}t' \: 
{\mathcal Y}_{t'} 
\right) \ ,
\label{qt}
\eea
where we defined
${\mathcal Y}_{t} = \{\PO X_{t} \PO + 
(1-\PO) X_{t} (1 - \PO)\}$.
Evidently this operator is block diagonal
in the space of chiral fields: ${\mathcal Y}_{t}= {\mathcal Y}_{t,L} + 
{\mathcal Y}_{t,R}$.
For closed gauge loops ($\Lambda_1 = \id$) the operator $Q_1$ is just
the path ordered exponential of the integral over ${\mathcal Y}_t$ and as
usual one finds
\bea
\det Q_1 = \exp \left( - \int_0^1 \mbox{d}t \: 
\mbox{Tr} \, {\mathcal Y}_t \right) \ .
\eea
The twist is just the determinant of $Q_1$ in the subspace of left-handed
fermion fields at $t=0$ and thus given as
\bea
\MT = \exp \left(- \int_0^1 \mbox{d}t \: 
\mbox{Tr} \, P_0 {\mathcal Y}_t \right) \ .
\label{T_trace}
\eea
The projector restricts the trace to the space of left-handed fields
only. With the definition \pref{def_cov_anomaly} for the covariant anomaly,
equation \pref{T_trace} can be rewritten as
\bea
\MT = \exp \left( i a^4 \sum_x {\mathcal A}^c_{t=0} (x) 
\int_0^1 \: \mbox{d}t \: X^c_t (x) \right)  \ .
\label{twist_gauge}
\eea
That is, along closed gauge loops the twist is
proportional to the exponential of the covariant anomaly
for the starting configuration.

Our derivation of this result assumes the gauge loop to be
smooth. This is not the case for $-\Gamma_3 \circ \Gamma_1$ we are
interested in. However,
eq. \pref{qt} can be applied independently to $\Gamma_1$ and
$\Gamma_3$. Taking the product of the two solutions for $Q$ we arrive
again at  \pref{twist_gauge}.

Now we notice that
$-\Gamma_3 \circ \Gamma_1$ starts in the vacuum configuration.
There the anomaly is a translational invariant field,
i.e. ${\cal A}^c(x) = {\cal A}^c(-x)$\footnote{recall the periodic
boundary conditions}.
Moreover, it is a pseudo-scalar field and therefore odd under parity:
${\cal A}^c(x) =-{\cal A}^c(-x)$. 
This implies that the anomaly vanishes in the vacuum of $SU(N)$.
An immediate consequence is the result
\be\label{twist_13}
\MT(-\Gamma_3 \circ \Gamma_1)=1\,.
\ee
Along our closed gauge loop the twist is trivial.

%
%

\vspace*{0.5cm}
\noindent {\bf{\slshape{4.3 The twist along the non-gauge loop 
${\mathbf \MT \, (\Gamma_2 \circ \Gamma_3)}$ }}}
\vspace*{0.5cm}

\noindent We are now left with the calculation of the twist for the
loop $\Gamma_2 \circ \Gamma_3$. This presents the added difficulty
of containing a non-gauge part, the path $\Gamma_2$. Instead of solving
the differential equation for $Q_t$  we will derive an integral
representation for the 
twist which we will use  for the computation of $\MT(\Gamma_2 \circ \Gamma_3)$.

So far we considered loops in configuration space, defined by some
gauge field $U_t$, depending on the parameter $t$. In the following we
assume the gauge field to be smoothly dependent on two parameters, $t$
and $s$, both lying in the range $[0,1]$.
In addition, the field $U_{t,s}$ should have the following two properties:
\bea\label{homotopy1} 
1)\:\: U_{0,s} & =& U_{1,s}\,=\,\id \,\quad 0\leq s \leq 1\,, \\ 
2)\:\: U_{t,0} &  \equiv & \id\,,\qquad \qquad 0\leq t \leq
1\,.\label{homotopy2} 
\eea
The first one tells us that $t$ parameterizes closed loops 
with constant base point $\id$ for all
allowed values of $s$. In the following we will simply write
$\Gamma(s)$ instead of $\Gamma(U_{t,s})$. One can think of $s$ as an
deformation parameter 
defining a homotopy between the two loops $\Gamma(0)$ and
$\Gamma(1)$.
The second property means that $\Gamma(0)$ is a constant
loop with a single configuration, the classical vacuum.

Now that the gauge field depends on two parameters, the same is true
for the projector $\hat{P}_- = P_{t,s}$ and the
evolution operator $Q=Q_{t,s}$.
Consider the twist for the loop $\Gamma(s)$. Starting from the
definition of the twist one can easily verify the relation
\bea
\dpar{s} \ln \MT (\Gamma(s)) = \Tr \, ({P_{0,s} \, Q_{1,s}^{-1}}
  \, \dpar{s} Q_{1,s})\ .
\label{ds}
\eea
Since $P_{0,s}=P_{1,s}$ 
the projector $P_{0,s}$ commutes with $Q_{1,s}$.
For the derivative of $Q_{t,s}$ with respect to $s$ 
an integral representation can be found. Let us make the ansatz
$\dpar{s} Q_{t,s} =Q_{t,s}R_{t,s}$. If we differentiate this equation with
respect to $t$ and the differential equation \pref{DE} with respect to
$s$, we  find 
\bea
\dpar{s}Q_{t,s} = Q_{t,s} \int_0^t \mbox{d}r \: 
Q_{r,s}^{-1} \dpar{s} X_{r,s} \, Q_{r,s} \ .
\label{RESULT}
\eea
Here we defined $X_{r,s}= [\dpar{r} P_{r,s},P_{r,s}]$.
Setting $t$ equal to 1 eq. \pref{ds} can alternatively be written as 
\bea
\dpar{s} \ln \MT (\Gamma(s)) =  \int_0^1 \mbox{d}t \, \Tr  P_{t,s} \, 
\, [\dpar{t} P_{t,s}, \dpar{s}P_{t,s}]\ .
\label{DS}
\eea
If we integrate this equation over $s$ we finally find an integral
representation for 
the twist along $\Gamma(1)$,
\bea\label{integralrepresentation}
\MT(\Gamma(1)) =   \exp \int_0^1\int_0^1\mbox{d}t\,\mbox{d}s \,
\Tr  P_{t,s} \,  
\, [\dpar{t} P_{t,s}, \dpar{s}P_{t,s}]\ .
\eea
It is only here that \pref{homotopy2} enters the result in terms of
$\MT(\Gamma(0)) = 1$.
One should keep in mind that
the integral representation \pref{integralrepresentation} is valid
provided the loop $\Gamma(1)$ can be smoothly contracted to
$\Gamma(0)$. This might not be the case for all possible
loops in the space of admissible gauge fields. The bound \pref{bound_plaquette}
is responsible for a highly non-trivial topology of that space.

Formula \pref{integralrepresentation} looks not very accessible. In
fact, the twist along $\Gamma(1)$ is expressed as an integral
over the surface defined by $U_{t,s}$. Notice that the
right-hand side is independent of the particular parameterization of
the surface.
However, the trace in \pref{integralrepresentation} can be expanded in
powers of the lattice spacing $a$. This is sufficient to compute the
twist for small $a$.

Consider $U_{t,s}$ to be a homotopy of classical fields. According to
\pref{classical_field} the link field $U_{t,s}$ is given as the
path-ordered exponential of the gauge potential $A_{\mu}(t,s,x)$,
depending also on the parameters $t$ and $s$. The expansion of the
trace was already discussed in ref. \cite{NonAbelianChGT}. There the
leading term is found to be
\be\label{expansion_trace}
\Tr  P_{t,s} [\dpar{t} P_{t,s}, \dpar{s}P_{t,s}]=
-ic_2\int\mbox{d}^4x\, d^{abc}\epsilon_{\mu\nu\rho\sigma}
\dpar{t}A_{\mu}^a(t,s,x)\dpar{s}A_{\nu}^b(t,s,x)
F_{\rho\sigma}^c(t,s,x) \ ,
\ee
where $F_{\rho\sigma}^c(t,s,x)$ denotes the
field tensor associated with the gauge potential. The constant $c_2$
equals $1/32\pi^2$ and $d^{abc}$ is the completely symmetric d-symbol
of the gauge group, defined by
\bea
d^{abc} = 2i \tr \{ T^a [T^b T^c + T^c T^b] \} \ .
\label{dsymbol}
\eea
If we use expansion \pref{expansion_trace} into
\pref{integralrepresentation} we are in principle able to compute the
twist for small $a$.

Let us now turn to the loop $\Gamma_2\circ\Gamma_3$ we are interested
in. All we need is a parameterization of the surface enclosed by it. 
Even less we only need the parameterization in the classical continuum
limit. A simple example is given by
\bea
A_{\mu}(t,s,x) = (1-t) \Omega(s,x) \partial_{\mu} \Omega(s,x)^{-1}\,,
\label{potential}
\eea
with $\Omega(s,x)$ as introduced before. This surface has the correct
boundary. For $t$ equal to 0 we recover the correct continuum limit of
$\Gamma_3$, defined in \pref{def_gamma_3}. On the other hand, setting $s$
equal to 1, eq. \pref{potential} coincides with \pref{path} for the
classical vacuum configuration, the continuum limit of $\Gamma_2$. 
Note that the parameters $t$ and $s$ do not coincide with the ones given
in \pref{homotopy1} and \pref{homotopy2}. 
We instead made use of the parameterization invariance
of the integral in \pref{integralrepresentation} to define a surface matching 
our particular boundary paths in the classical continuum limit.

The surface parameterization leads directly to the following
relations, needed in \pref{expansion_trace}: 
\be
\begin{array}{rcl}
\partial_s A_{\mu}(t,s,x)  &=& (1-t)
\left\{[\Omega\partial_{\mu}\Omega^{-1},
\Omega\partial_{s}\Omega^{-1}]
+ \partial_{\mu}
(\Omega\partial_{s}\Omega^{-1})
\right\} \ , \\[2ex]
\partial_t A_{\nu}(t,s,x)& = &-
\Omega\partial_{\nu}\Omega^{-1} \ , \\[2ex]
F_{\rho \sigma}(t,s,x)& =& (t^2 - t) \,
[\Omega\partial_{\rho}\Omega^{-1},
\Omega\partial_{\sigma}\Omega^{-1}] \ .
\label{defor}
\end{array}
\ee
For brevity we have suppressed the dependence on $(s,x)$ in $\Omega$.
The integration over $t$ in \pref{integralrepresentation} is easily
performed and gives a factor $1/12$. Making use of the
anti-symmetry property of the $\epsilon$-tensor we finally find
\be
\ln \MT (\Gamma_2 \circ \Gamma_3) 
= \frac{-1}{48\pi^2} \int\limits_0^1\mbox{d}s 
\int\mbox{d}^4x\,
\epsilon_{\mu \nu \rho \sigma}
\tr \left(\gdg{s}\gdg{\mu} \cdots  \gdg{\sigma}\right)
\label{step}
\ee
for the leading term in the expansion in $a$.
Note that we multiplied back the derivatives of the potential and the
field tensor into the trace over the group generators. Keep also in
mind that they generate the fundamental representation of $SU(3)$.

To bring the integral into a more familiar form we define
$x_4= s$ and introduce a five dimensional tensor 
$\epsilon_{\mu \nu \rho \sigma \lambda}$ with 
$\epsilon_{01234} = 1$. 
The logarithm of the twist can thus be written as
\be
\ln \MT (\Gamma_2 \circ \Gamma_3) 
= \frac{-1}{240 \pi^2} \int\limits_{0 \leq x_4 \leq 1} \mbox{d}^5 x \, 
\epsilon_{\mu \nu \rho \sigma \lambda}
\tr(\gdg{\mu} \cdots \gdg{\lambda}) .
\label{final}
\ee
A factor $1/5$ compensates the extra terms
due to the fifth index of the anti-symmetric tensor.

This type of integrals is well known in the context of
anomalies in the continuum. 
Usually one considers mappings $\Omega(s,x)$ with boundary values
$\Omega(0,x) =\Omega(1,x)=\id$. In that case the
integral is an integer multiple of $2\pi i$. In \pref{final}, however, the
boundary values are given by \pref{SU3_path}.
Even in this case the possible values are strongly restricted.
First note that the integral is invariant under small variations 
\be
\delta\Omega(s,x) \,=\,\omega^c(s,x)T^c\,,
\ee
as long as the variation preserves
the particular form of the boundary values \pref{SU3_path}.
Therefore, the integral depends only on the homotopy class of
$g(x)$. If $g(x)$ is in the non-trivial homotopy class, $g^2(x)$
belongs to the trivial one. This already implies that the integral in
\pref{final} is either $0$ or $i\pi$ (mod $2\pi i$). 

An explicit computation, performed by Witten in \cite{Witten2}, yielded the
value $i\pi$. 
Hence our final result is
\be\label{twist_23}
\MT(\Gamma_2 \circ \Gamma_3)=-1\,
\ee
in the classical continuum limit.

\vspace*{0.5cm}
\noindent {\bf {\slshape {4.4 Global anomaly in SU(2)}}}
\vspace*{0.5cm}

\noindent Taking into account result \pref{twist_13} 
the composition law \pref{composition_law} reads
\be\label{eff_comp_law}
\MT(\Gamma_2 \circ \Gamma_1)=\MT(\Gamma_2 \circ \Gamma_3).
\ee
As we have already mentioned the left hand side of this equation can
only take the 
values $\pm 1$. Since \pref{eff_comp_law} is valid for all lattice
spacings, it implies that \pref{twist_23} is not only the result
in the continuum limit, but it is also the twist
for a sufficiently small lattice spacing $a$. The absence of terms
proportional to $a$ in \pref{twist_23} has its reason in our special
loop which starts in the vacuum configuration. So our final result
reads 
\be\label{result_twist_12}
\MT(\Gamma_2 \circ \Gamma_1)=-1.
\ee
We want to emphasize that the introduction of $\Gamma_3$ is merely a
technical tool. If we were able to compute directly the evolution
operator $Q_t$ along $\Gamma_2 \circ \Gamma_1$, we would have found
the same twist using directly its definition \pref{def_twist}.

We made use of the fact that 
$\Gamma_2 \circ \Gamma_1$ starts in the classical vacuum
configuration.
Our result \pref{result_twist_12}, however, is not restricted to such loops.
As we have previously remarked, the twist is a homotopy invariant
in $SU(2)$. Having found a loop with a twist equal to $-1$,
all loops in the same homotopy class have the same twist.
This holds true even though the loop
we have considered is only piecewise
smooth, since it can be deformed to a totally
smooth one.

To establish a global anomaly we have to show $W\neq\MT$ along 
$\Gamma_2 \circ \Gamma_1$. The semi-classical argument in section 3.2
cannot directly be applied to this loop, because $\Gamma_1$ has no
continuum limit. However, the current is of order $a$ along
$\Gamma_2$. The path $\Gamma_1$ is a curve of gauge transformations
and because the current is gauge-covariant we can conclude that the
current is of order $a$ along the entire loop. Hence for the
Wilson line along $\Gamma_2 \circ \Gamma_1$ we find
\be
W(\Gamma_2 \circ \Gamma_1)\,=\,1+{\cal O}(a).
\ee
According to our discussion in section 3, the loop
$\Gamma_2 \circ \Gamma_1$ thus violates the integrability condition and an
$SU(2)$ gauge theory exhibits a global anomaly.

As we have already pointed out, the integral in \pref{final} 
was already encountered in the context of the Witten anomaly.
In ref. \cite{Embedding} the Witten anomaly is calculated as 
the cumulative effect of the perturbative non-abelian 
anomaly along a path representing the
continuum version of $\Gamma_3$.
The change in the phase of the effective action 
under homotopically non-trivial $SU(2)$ gauge transformations
is precisely given by the integral in (\ref{final}).
This already indicates a connection between Witten's anomaly and the
obstruction we found. In section 6 we will have a closer look at
this.\footnote{The reader mainly interested in the
  connection to Witten's anomaly 
  can safely skip the next section and may continue directly 
  with section 6 in a first reading.}

%
%

\setcounter{equation}{0}
\renewcommand{\theequation}{5.\arabic{equation}}

\vspace*{0.5cm}
\noindent{\large{{\bf{{\slshape{5. Global SU(2) anomaly in higher
            representations}
}}}}}
\vspace*{0.5cm}

\noindent{{\bf{{\slshape{5.1 Preliminaries} 
}}}}
\vspace*{0.5cm}

\noindent So far we considered the fundamental representation of
$SU(2)$ only. For computational reasons we introduced an auxiliary
path $\Gamma_3$ with the gauge field in the fundamental representation
of $SU(3)$ and embedded the links of the other paths $\Gamma_1,\Gamma_2$,  
into this representation.

In the following we compute the twist for
higher representations of $SU(2)$ along the same lines. However,  the embedding
turns out to be less 
trivial. Basically, for any given $SU(2)$ representation $j$ we cannot
find an appropriate $SU(3)$ representation $R$ that will decompose into
$j$ and a finite number of singlets.
More generally, we always  
end up with a decomposition
\bea\label{SU2_rep_decomposition}
R &\rightarrow& \sum_j c(R,j)\,j\,,
\eea
where the coefficients $c(R,j)$ give the multiplicity of the
representation $j$. To turn \pref{SU2_rep_decomposition} the other way
around, only particular sums of $SU(2)$ representations can be
embedded in an $SU(3)$ representation $R$, such that a path 
of gauge transformations \pref{SU3_path} exists.

Having this in mind we start with the definition of
$\Gamma_3$. In analogy to \pref{def_gamma_3} we define $\Gamma_3(R)$
with a curve 
$\Omega_R(s,x)$, this time in the representation $R$ of $SU(3)$. The
boundary value $\Omega_R(1,x)$ is still in $SU(2)$, but now in the
representation 
given by the right hand side of
\pref{SU2_rep_decomposition}. Consequently we define the paths
$\Gamma_1,\Gamma_2$ for this direct sum of $SU(2)$ representations.

Using \pref{direct_sum_twist} several times we find the generalization
of \pref{compo} to be
\bea\label{master_twist}
\prod_j\twist(j)^{c(R,j)} & = & \twist(R)\,. 
\eea
Notice that $\twist(R)$ denotes the twist along
$\Gamma_2\circ\Gamma_3$ and $R$ indicates the representation used for
$\Gamma_3$. 
On the other hand, $\twist(j)$ stands for the twist along
the pure $SU(2)$ loop $\Gamma_2\circ\Gamma_1$. In \pref{master_twist}
we already used the fact that the twist along the pure gauge loop
$-\Gamma_3\circ\Gamma_1$ is 1 for all representations $R$.
Relation \pref{master_twist} is our master formula that will enable us
to compute recursively \twist($j$) for all $SU(2)$ representations.

\vspace*{0.5cm}
\noindent{{\bf{{\slshape{5.2 The twist ${\mathbf \twist(R)}$} 
}}}}
\vspace*{0.5cm}

\noindent 
Let us discuss the twist $\twist(R)$ along
$\Gamma_2\circ\Gamma_3$ for arbitrary $R$.
The integral in \pref{final} contains the product of five
generators of $SU(3)$. Expressing twice the product of two of them as a
commutator, we can write our result \pref{final} as
\bea\label{ln_twist_fund}
\ln\twist\ &=& \dsym I^{abc},
\eea
where \dsym\ is the d-symbol of $SU(3)$.
Notice that \dsym\ is real because of the anti-hermiticity of the
generators. The remaining part $I^{abc}$ is independent of the group
representation and the explicit expression of it is not needed in the
following. 
For some other representation $R$ the result \pref{final}
can be similarly  written as
\bea
\ln\twist(R) &=& \dsymR I^{abc}.
\eea  
It differs from \pref{ln_twist_fund} only in the $d$-symbol that is
now defined with the generators $R(T^a)$ of the representation $R$.
One can show that \dsymR can always be expressed in terms of \dsym.
Introducing the {\em anomaly coefficient} $A(R)$ we write
\bea\label{def_A_coeff}
\dsymR &=& A(R)\,\dsym\,,
\eea
leading to the relation
\bea
\twist(R) & = & \twist^{A(R)}.
\eea
for the twist along $\Gamma_2\circ\Gamma_3$ in higher
representations. Once the twist \twist\ for 
the fundamental representation is known, 
its value for higher representations is completely determined by the
anomaly coefficient. The computation of \twist$(R)$ is therefore
reduced to the purely group theoretic task of calculating $A(R)$. 

By definition $A(R)$ is equal to one for the fundamental
representation. Furthermore one can straightforwardly establish the relations
\be\label{relations_anomaly_coeff}
\ba{rcl}
A(R_1\oplus R_2) & = & A(R_1)+A(R_2),\\
A(R_1\otimes R_2) & = & A(R_1)d_2+A(R_2)d_1,
\ea
\ee
where $d_i$ denotes the dimension of the representation $R_i$. 

Let us consider a particular example. The
fundamental representation of $SU(3)$ 
is three dimensional and in the following denoted by $3$. For the n-fold tensor
product $3\otimes\ldots\otimes3=3\otimes^n$ the anomaly coefficient 
and the twist are
easily computed to be 
\be\label{twist_3_n}
A(3\otimes^n) \, = \, n3^{n-1},\qquad\twist(3\otimes^n) \,=
\, (-1)^n . 
\ee
We will use this result in the next subsection to classify the anomalous
$SU(2)$ representations.

\vspace*{0.5cm}
\noindent{{\bf{{\slshape{5.3 Global SU(2) anomaly in higher
          representations}  
}}}}
\vspace*{0.5cm}

\noindent As already mentioned, the fundamental representation 3 of $SU(3)$ 
decomposes into a doublet and a singlet of $SU(2)$. Hence, for the
n-fold tensor product $3\otimes^n$ the decomposition
\pref{SU2_rep_decomposition} reads 
\bea\label{decomp_su2}
3\otimes^n &\rightarrow&
\big(\frac{1}{2}\,\oplus\,0\big)\otimes^n\quad=\quad\sum_j
c(n,j)\,j\,. 
\eea
Up to $n$ equals 5 we have collected the explicit decompositions in
table \ref{table1}. Consider first $n=2$.
\begin{table}[t]
\bea
\big(\frac{1}{2}\oplus 0\big)\otimes^2 & = &\phantom{1}2\cdot 0\,
\oplus\phantom{1}
2\cdot\frac{1}{2}\,\oplus\phantom{2}1\cdot1\nn\\ 
\big(\frac{1}{2}\oplus 0\big)\otimes^3 & = &\phantom{1}4\cdot
0\,\oplus\phantom{1} 5\cdot\frac{1}{2}\,\oplus\phantom{2} 3\cdot
1\,\oplus\phantom{1}1\cdot\frac{3}{2}\nn\\ 
\big(\frac{1}{2}\oplus 0\big)\otimes^4 & = &\phantom{1}9\cdot
0\,\oplus 12\cdot\frac{1}{2}\,\oplus\phantom{1} 9\cdot
1\,\oplus\phantom{1} 4\cdot\frac{3}{2}\,\oplus1\cdot 2\nn\\ 
\big(\frac{1}{2}\oplus 0\big)\otimes^5 & = &21\cdot
0\,\oplus 30\cdot\frac{1}{2}\,\oplus 25\cdot 1\,\oplus
14\cdot\frac{3}{2}\,\oplus 
5\,\cdot2\,\oplus1\cdot\frac{5}{2}\nn
\eea
\caption{The decomposition of $\big(\frac{1}{2}\oplus 0\big)\otimes^n$ up to
  $n$ equals 5.}
\label{table1}
\end{table}
Using \pref{master_twist} and \pref{twist_3_n} we find
\bea
\twist(1) &= & 1\,,
\eea
because the anomalous doublet representation with
$\twist(\frac{1}{2})=-1$ appears twice in the
decomposition. On the other hand,
$c(3,\frac{1}{2})$ 
equals 5 and we find 
\bea
\twist(\frac{3}{2})\twist(\frac{1}{2}) \,=\,-1 & \Rightarrow&
\twist(\frac{3}{2})\,=\, 1\,.
\eea
Here the minus sign coming from \pref{twist_3_n} is compensated by an
additional sign due to an odd number of doublet representations. 
Compare this with $n=5$. Here $c(5,\frac{1}{2})$ is
even and, because one can check before $\twist(2)=1$, we get
\be
\twist(\frac{5}{2}) \, = \, -1\,.
\ee
These explicit examples should be enough to illustrate the idea of our
method. In general we find the following results for all integers
$n$ including zero:
\be
\ba{llcr}
\mbox{1.} & \twist(n)  &= & 1,\\[0.3ex]
\mbox{2.} & \twist(2n+\frac{1}{2})& = & -1,\\[0.3ex]
\mbox{3.} & \twist(2n+\frac{3}{2})& = & 1.
\ea
\ee
The proofs for these statements can be found in appendix A.
In summary we find the representations with
\be\label{result_twist_general}
j \, = \, 2n+\frac{1}{2}\,,\qquad n\,=\,0,1,2\ldots
\ee
to be anomalous.

\setcounter{equation}{0}
\renewcommand{\theequation}{6.\arabic{equation}}

\vspace*{0.5cm}
\noindent{\large{{\bf{{\slshape{6. Connection with Witten's
anomaly\label{page_Connection_Witten}
}}}}}}
\vspace*{0.5cm}

\noindent
In the following we will show that the global anomaly we discovered 
in the previous sections is nothing but the Witten anomaly \cite{Witten}.
Let us consider a smooth closed loop $U_t$ in field space as
above. For simplicity we assume $D_0$ to have no zero modes.
The function
\bea
f(t)  =  \det \, (1-P_+ + P_+D_tQ_tD_0^{\dagger})
\eea
depends smoothly on $t$ and is
real for the gauge group $SU(2)$. In addition it 
satisfies
\bea
f(0)\,>\,0\,, & & f(1) = {\cal T}f(0).
\eea
This implies  that $f$
passes through zero an odd number of times for $0\leq t\leq 1$ if and
only if ${\cal T}=-1$.
Next one can show
\bea\label{f_squared}
f^2 (t) =  \det D_t \det D_0^{\dagger}.
\eea
Because of the $\gamma_5$-hermiticity $D^{\dagger}=\gamma_5 D
\gamma_5$, the eigenvalues $\lambda_i$ of $D$  
come in complex conjugate pairs and for the determinant we find
\bea
\det D_t=\prod_i\lambda_i(t)\lambda_i^*(t).
\eea
The eigenvalues depend smoothly on $t$ for smooth paths.
According to \pref{f_squared} a
passing through zero of $f(t)$ at some point $t_0$ implies a passing
through zero 
of an odd number of
eigenvalues $\lambda_i(t)$. One can prove this by expanding
both $f$ and $\lambda_i$ around $t_0$ in \pref{f_squared}.
Thus we conclude that an odd number of eigenvalues of $D$ cross zero
along a closed loop if and only if $\MT =-1$.

We gave an explicit example for such a loop,
namely $\Gamma_2\circ\Gamma_1$. Because $\Gamma_1$ is a curve of gauge
transformations the crossing of the eigenvalues occurs along
$\Gamma_2$, the lattice analogue of \pref{path}. 
Thus we find the same behaviour of the spectral flow on the lattice
that Witten proved in the continuum using the Atiyah--Singer index theorem.

Witten also investigated the anomaly for
higher fermion representations. He found the theory to be
inconsistent if the representation is such that $2\,\mbox{tr} \,T_3^2$
is an odd integer.
The trace of $T_3^2$ depends only
on $j$ and one finds
\bea\label{anomaly_cond_2}
2\,\mbox{tr}\, T_3^2 & = & \frac{2}{3}j(j+1)(2j+1).
\eea
Obviously the right hand side of \pref{anomaly_cond_2} is even if $j$ is an
integer, so these representations are anomaly free. For half-integer
values one easily proves that $2\,\mbox{tr}\, T_3^2$ is odd if
\bea\label{anomaly_cond_3} 
j&=&2n+\frac{1}{2}\,,\qquad n=0,1,2,\ldots
\eea
As has been shown in the previous section,
these are exactly the representations with a twist equal to $-1$ on the
lattice. Notice that this coincidence is highly non-trivial.
Equation \pref{anomaly_cond_3} emerges from the investigation of the
zero modes of a five dimensional Dirac operator in a certain instanton
field. This differs completely from the embedding technique that leads
to \pref{result_twist_general}.

\setcounter{equation}{0}
\renewcommand{\theequation}{7.\arabic{equation}}

%
%

\vspace*{0.5cm}
\noindent {\large {{\bf {{\slshape {7. Concluding remarks}}}}}}
\vspace*{0.5cm}

\noindent In this paper we have addressed
for the first time the issue of non-perturbative
anomalies beyond the semi-classical level. The
lattice formulation provides a mathematically well-defined
framework to analyze this problem. 

Global anomalies on the lattice are
related to a non-trivial twist 
for closed loops in configuration space. The
twist $\MT$ has proved to be the essential quantity.
In particular we did not consider
directly the
effective action to establish the anomaly.
The twist has the advantage to be also defined in sectors
with non-zero topological charge and might be considered to
investigate the existence of global anomalies there.

An explicit construction of a non-trivial closed loop in the vacuum sector 
has been given for the case of an $SU(2)$ gauge theory coupled
to a doublet of Weyl fermions. 
More generally, the anomaly is present
whenever we consider representations of $SU(2)$ such 
that $2\, \mbox{tr}\, T_3^2$ is an odd integer.

Our treatment of global anomalies has the advantage of being 
completely analytical
and resembles rather  the embedding technique 
used in \cite{Embedding}
than the original proof of Witten, based on the spectral flow of the
Dirac operator. However, we have shown
that Witten's original argument follows immediately from 
our result.

In L\"uscher's approach only those chiral gauge 
theories can be formulated consistently on the lattice which are 
perturbatively anomaly free in the continuum
\cite{AbelianChGT,NonAbelianChGT}.
This means that a necessary condition for the lattice 
formulation to exist is an anomaly free fermion
multiplet ($d_R^{abc} = 0$).
We have seen here that this is also the case for non-perturbative
anomalies as far as Witten's anomaly is concerned.
However, one needs to exclude other global anomalies.
Starting from a fermion representation perturbatively
anomaly free it is not obvious that
$\MT$ will be equal to one for all closed loops. Only in that
case one can set $j_{\mu}^c = 0$. The four conditions listed in section
3 are then satisfied and the chiral gauge theory is completely defined.

\vspace*{0.5cm}

We want to thank Martin L\"uscher
for  advise and numerous comments throughout the realization
of this work. Thanks go also to Peter Weisz 
who critically read the first version of this paper.

\newpage
%
%
\subsubsection*{Appendix A}
\setcounter{equation}{0}
\renewcommand{\theequation}{A.\arabic{equation}}
%

{\bf I.} $\twist(n) = 1$ for all integers $n$ including zero.\\[0.6ex]
\noindent {\em Proof:} The statement is true for $n=0,1$. Now we assume 
$\twist(k)=1$ for $k=1,2,\ldots,n$. Consider
\be
(\frac{1}{2}\oplus0)\otimes^{2n+2}\,=\,(1\oplus2\cdot\frac{1}{2} \oplus
2\cdot0)\otimes^{n+1} \,=\, 1\otimes^{n+1}\,+\,\sum_j \tilde{c}(2n+2,j)j\,.
\ee
Because both the doublet and the singlet representation appear twice in the
decomposition of $(1/2\oplus0)\otimes^2$, the coefficients
$\tilde{c}(2n+2,j)$ are 
even for all values of $j$. The tensor product of the triplet
representation can be decomposed further and we write
\be
1\otimes^{n+1}\,=\,(n+1) \oplus \sum_{k=0}^n a_k\,k\,,
\ee
where we introduced the not necessarily even integers $a_k$. In that
decomposition only integer representations occur. Taking into account
\pref{twist_3_n} equation \pref{master_twist} reads
\be
\twist(n+1)\prod_k\twist(k)^{a_k}\prod_j\twist(j)^{\tilde{c}(2n+2,j)}\,=\,
1\,.
\ee
According to our assumption $\twist(k)=1$ this implies
$\twist(n+1)=1$.\hfill$\newsquare$\\

\noindent{\bf II.} $\twist(2n+\frac{1}{2}) = -1$ and
$\twist(2n+\frac{3}{2}) = 1$ for all integers $n$ including zero.\\[0.6ex]
\noindent {\em Proof:} The statement is true for $n=0$. To prove it for
arbitrary integers we first define
\bea
\Sigma_a(2n+\frac{1}{2}) & =
&\sum_{k=0}^{n-1}c(4n+1,\frac{4k+1}{2}),\qquad n\geq 1\,,\\
\Sigma_a(2n+\frac{3}{2}) & =
&\sum_{k=0}^{n}c(4n+3,\frac{4k+1}{2}),\qquad n\geq 0\,.
\eea
$\Sigma_a(j)$ gives the total number of anomalous representations in
the decomposition \pref{SU2_rep_decomposition} of $(\frac{1}{2}\oplus
0)\otimes^{2j}$ with 
highest weight less than $j$. In addition we define 
\bea
\Sigma_f(2n+\frac{1}{2}) & =
&\sum_{k=1}^{n}c(4n+1,\frac{4k-1}{2}),\qquad n\geq 1\,,\\
\Sigma_f(2n+\frac{3}{2}) & =
&\sum_{k=1}^{n}c(4n+3,\frac{4k-1}{2}),\qquad n\geq 1\,.
\eea
$\Sigma_f(j)$ gives the number of anomaly free {\em half integer}
representations in 
the decomposition of $(\frac{1}{2}\oplus 0)\otimes^{2j}$, again with
highest weight less than $j$.
Formulae \pref{twist_3_n} and \pref{master_twist} imply 
\bea
\twist(j)\cdot (-1)^{\sum_a(j)} & = & -1
\eea
for half-integer $j$.
In order to prove the statement all we need to show is that
$\Sigma_a(2n+\frac{1}{2})$ is even and $\Sigma_a(2n+\frac{3}{2})$ is
odd. 

Using table \pref{table1} one can easily check that $\Sigma_a(5/2)$ is
even and $\Sigma_a(3/2)$ odd. 
Now let us assume that 
$\Sigma_a(2n+\frac{1}{2})$ is even for some fixed integer $n\geq 1$.
We consider the tensor product 
\bea\label{tensor_prod_general}
(\frac{1}{2}\oplus0)\otimes^{4n+3} & = & 
1\otimes
\left((\frac{1}{2}\oplus0)\otimes^{4n+1}\right)\, \oplus
  \sum_j\tilde{c}(4n+3,j)j\,,
\eea 
where all the coefficients $\tilde{c}$ are even integers. The tensor
product with the triplet representation can be decomposed into
irreducible representations with the result
\be\label{decomp_int_general}
1\otimes
\left((\frac{1}{2}\oplus0)\otimes^{4n+1}\right)\, =\,
\sum_{k=1}^{2n+2}b(4n+3,\frac{2k-1}{2})\frac{2k-1}{2}\, \oplus\,
\mbox{integer representations}\,.
\ee
The coefficients $b$ are given in terms of the coefficients $c$ as 
\be\label{general_b}
\ba{lclclcl}
b(4n+3,\frac{1}{2})& = & c(4n+1,\frac{1}{2}) & + &
c(4n+1,\frac{3}{2}) ,& \\[0.8ex]
\hspace{1.15cm}\vdots & &\hspace{1.15cm}\vdots &
&\hspace{1.15cm}\vdots & \\[0.8ex] 
b(4n+3,\frac{2k+1}{2}) & = & c(4n+1,\frac{2k-1}{2}) & + &
c(4n+1,\frac{2k+1}{2})& + & c(4n+1,\frac{2k+3}{2}), \\[0.8ex]
\hspace{1.15cm}\vdots & &\hspace{1.15cm}\vdots &
&\hspace{1.15cm}\vdots & & \hspace{1.15cm}\vdots\\[0.8ex] 
b(4n+3,\frac{4n+1}{2})& =  & & & c(4n+1,\frac{4n-3}{2}) & + &
c(4n+1,\frac{4n-1}{2}), \\[0.8ex]
b(4n+3,\frac{4n+3}{2})& = & & & & &
c(4n+1,\frac{4n-1}{2}).\\[2ex]
\ea
\ee
\vspace{0.3cm}

\noindent We denote the contribution of \pref{decomp_int_general} to
$\Sigma_a(2n+\frac{3}{2})$ by
$\tilde{\Sigma}_a(2n+\frac{3}{2})$. Using $c(m,\frac{m}{2})=1$ we find
\bea\label{result_sigma_tilde}
\tilde{\Sigma}_a(2n+\frac{3}{2})& = & 1 + \Sigma_a(2n+\frac{1}{2}) +
2\Sigma_f(2n+\frac{1}{2})\,.
\eea
We assumed $\Sigma_a(2n+\frac{1}{2})$ to be even, hence
$\tilde{\Sigma}_a(2n+\frac{3}{2})$ is 
odd. Because the coefficients $\tilde{c}$ in
\pref{tensor_prod_general} are all even, 
$\Sigma_a(2n+\frac{3}{2})$ is also odd, i.e.
\bea
\Sigma_a(2n+\frac{1}{2}) \mbox{ is even} & \Longrightarrow &
\Sigma_a(2n+\frac{3}{2}) \mbox{ is odd}\,.
\eea 
Having established this
result, we consider now the tensor product \pref{tensor_prod_general}
with exponent $4n+5$. Equations \pref{tensor_prod_general},
\pref{decomp_int_general} and \pref{general_b} need just be modified
by the replacement $n\rightarrow n+1/2$. Similar to
\pref{result_sigma_tilde} we find
\bea
\tilde{\Sigma}_a(2n+\frac{5}{2}) & = & 1 +
\Sigma_a(2n+\frac{3}{2})+2\,\Sigma_f(2n+\frac{3}{2})\,.
\eea
We therefore conclude:
\bea
\Sigma_a(2n+\frac{3}{2}) \mbox{ is odd} & \Longrightarrow &
\Sigma_a(2(n+1)+\frac{1}{2}) \mbox{ is even}\,.
\eea
This completes the proof.
\hfill$\newsquare$

\end{document}